\documentstyle[prb,aps,twocolumn,epsfig]{revtex}
\begin{document}
\draft
\twocolumn[\hsize\textwidth\columnwidth\hsize\csname @twocolumnfalse\endcsname\title{Temperature-dependent Raman spectroscopy in BaRuO$_3$ systems}
\title{Temperature-dependent Raman spectroscopy in BaRuO$_3$ systems}
\author{Y. S. Lee\cite{email1}}
\address{Center for Strongly Correlated Materials Research, \\
Seoul National University, Seoul 151-747, Korea}
\author{T. W. Noh}
\address{School of Physics and Research Center for Oxide Electronics, \\
Seoul National University, Seoul 151-747, Korea}
\author{J. H. Park and K.-B. Lee}
\address{Department of Physics, Pohang University of Science and Technology, Pohang,\\
Korea}
\author{G. Cao\cite{address} and J. E. Crow}
\address{National High Magnetic Field Laboratory, Florida State University,\\
Tallahassee, Florida 32306}
\author{M. K. Lee and C. B. Eom}
\address{Department of Material Science and Engineering, University of\\
Wisconsin-Madison, Madison, Wisconsin 53706}
\author{E. J. Oh and In-Sang Yang\cite{email2}}
\address{Department of Physics, Ewha Womans University, Seoul 120-750, Korea}
\date{\today }
\maketitle

\begin{abstract}
We investigated the temperature-dependence of the Raman spectra of a
nine-layer BaRuO$_3$ single crystal and a four-layer BaRuO$_3$ epitaxial
film, which show pseudogap formations in their metallic states. From the
polarized and depolarized spectra, the observed phonon modes are assigned
properly according to the predictions of group theory analysis. In both
compounds, with decreasing temperature, while $A_{1g}$ modes show a strong
hardening, $E_g$ (or $E_{2g}$) modes experience a softening or no
significant shift. Their different temperature-dependent behaviors could be
related to a direct Ru metal-bonding through the face-sharing of RuO$_6$. It
is also observed that another $E_{2g}$ mode of the oxygen participating in
the face-sharing becomes split at low temperatures in the four layer BaRuO$_3
$. And, the temperature-dependence of the Raman continua between 250 $\sim $
600 cm$^{-1}$ is strongly correlated to the square of the plasma frequency.
Our observations imply that there should be a structural instability in the
face-shared structure, which could be closely related to the pseudogap
formation of BaRuO$_3$ systems.
\end{abstract}


\vskip1pc] \newpage
%\newpage

\section{Introduction}

Recently, it was reported that a pseudogap formation can occur in 4$d$
transition metal oxides, BaRuO$_3$ compounds.\cite{yslee1,yslee2} Optical
conductivity spectra $\sigma _1(\omega )$ of both four-layer hexagonal (4$H$%
) BaRuO$_3$ and nine-layer rhombohedral (9$R$) compounds show clear
electrodynamic response changes resulting from the pseudogap formation. In
their metallic states, concurrent developments of a gap-like feature and a
coherent mode below a gap-like feature are observed due to the partial-gap
opening in the Fermi surface.

The pseudogap formations in the ruthenates could be closely related to their
structures characterized by hexagonal close-packing. As shown in Fig. 1,
their layered structures include the face-sharing structure of RuO$_6$
octahedra along the $c$-axis. In 4$H$ and 9$R$ structures, two and three
adjacent RuO$_6$ octahedra participate in face-sharing, respectively. A
direct Ru-Ru metal-bonding formed through such face-sharing distinguishes
their physical properties from those of the perovskite ruthenates only with
the Ru-O-Ru interaction through corner-sharing.\cite{cava99} Actually, the
metal-bonding has been seen to be closely related to interesting physical
properties, such as a metal-insulator transition in Ti$_2$O$_3$\cite{Ti2O3}
and non-Fermi liquid behavior in La$_4$Ru$_9$O$_{16}$.\cite{Khalifah01} In
the case of BaRuO$_3$ systems, the quasi-one-dimensional (1D) Ru
metal-bonding along the $c$-axis might induce a charge density wave (CDW)
instability. It was reported that a 5$d$ transition metal oxide 9$R$ BaIrO$_3
$, which is expected to have stronger metal-bonding character than 9$R$ BaRuO%
$_3$ due to a more extended 5$d$-orbital character, shows a static CDW
instability.\cite{Cao00} For Ba$M$O$_3$ ($M$=Ru, Ir) with 9$R$ or 4$H$
structure, it was observed that the strength of the metal-bonding through
the face-sharing should be a parameter in determining their physical
properties.\cite{yslee2} These strongly indicate that a CDW instability
could be related to the pseudogap formation in the BaRuO$_3$ systems.

In usual 1D density wave systems, a static CDW ordering state accompanies a
structural distortion with a metal-insulator transition. When Peierls-type
lattice distortion occurs, in general, new phonon modes in infrared (IR) and
Raman spectra can be observed due to structural symmetry breaking. This
could also result in a superlattice or an additional peak in x-ray
diffraction (XRD) patterns or neutron scattering experiments. In the case of
BaRuO$_3$ systems, although a 1D-like CDW instability is strongly suggested,
there has been no structural report about structural distortions.\cite
{comment1} These might be related to a fluctuation-type instability without
any static CDW ordering.

\begin{figure}[tbp]
\epsfig{file=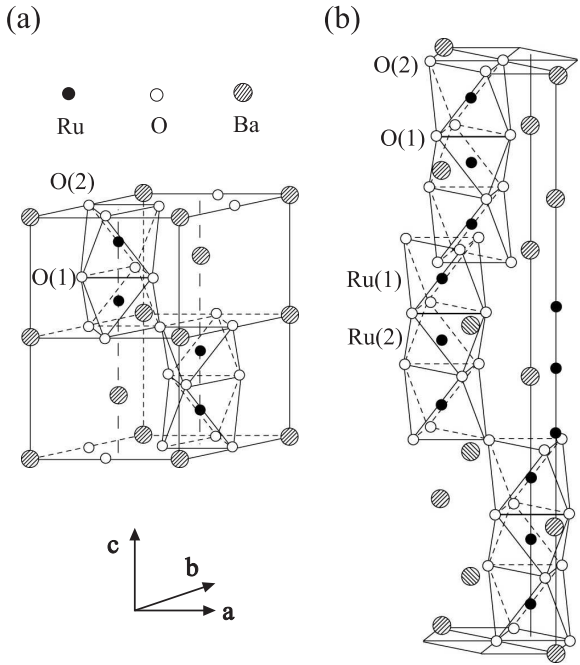,width=3.0in,clip=}
\vspace{2mm}
\caption{Schematic diagrams of the two crystallographic forms of BaRuO$_3$;
(a) 4$H$ phase, (b) 9$R$ phase. The arrows represent the crystallographic
axes. The details are described in the text.}
\label{structure}
\end{figure}

When the time scale of CDW fluctuations is long enough to induce a pseudogap
formation in $\sigma _1(\omega )$, a phonon anomaly, such as the creation or
splitting of a phonon, can be observed.\cite
{yslee3,Blumberg98,Lin98,Argyriou2000} In the metallic state of BaRuO$_3$
compounds, the screening of free carriers makes the detailed analysis of
IR-active phonons difficult. On the other hand, Raman spectroscopy is known
to be less affected by free-carrier responses than IR spectroscopy. So,
Raman spectroscopy could be a useful tool to address the origin of pseudogap
formation of BaRuO$_3$ compounds in view of their structural properties.

In this paper, we report the Raman spectra of 4$H$ and 9$R$ BaRuO$_3$
compounds. According to group theory analysis, the observed phonon modes are
properly assigned. From the temperature ($T$)-dependent experiments, it is
observed that the $T$-dependent behavior of phonons strongly depends on the
vibrational directions, which could be related to the structural
characteristics with the face-sharing of RuO$_6$ octahedra. Interestingly,
the $E_{2g}$ mode of the face-shared oxygen in 4$H$ BaRuO$_3$ becomes split
with decreasing $T$. These observations indicate that there should be a
structural instability due to the metal-bonding through the face-sharing of
RuO$_6$ octahedra, which could be closely related to the pseudogap formation
in these ruthenates.

\section{Experimentals}

4{\it H} BaRuO$_3$ epitaxial film on (111) SrTiO$_3$ substrate was
fabricated by a 90$^{\text{o}}$ off-axis sputtering technique. \cite{Eom00}
Its thickness is about 3200 \AA . XRD and transmission electron microscopy
reveal that the film is composed of a high quality single domain with a {\it %
c}-axis structure. 9{\it R} BaRuO$_3$ single crystal was prepared by a
flux-melting method.\cite{shepard97} The size of the sample is 0.5$\times $%
0.5$\times $0.2 mm$^3$. XRD measurements showed that the {\it c-}axis was
pointing along the short dimension. Due to the size limitation, most Raman
spectra were obtained in the {\it ab}-plane.

Raman scattering measurements were performed in the backscattering geometry
using a triple Raman spectrometer (Jobin Yvon T64000). The incident laser
beam was the 514.5 nm line of an Ar-ion laser and the laser power was about
3.6 mW on the sample surface. Raman spectra were measured at various $T$
between 5 K and 650 K. Below 300 K, a continuous flow type of cryostat was
used. Above 300 K, a home-made sample heating system was used. Due to the
heating effect of the focused laser, the assigned temperatures in this paper
could be slightly different from the actual ones on the measured sample
surface. At all $T$, polarized and depolarized spectra were obtained and
corrected by a Bose-Einstein factor. The details are described elsewhere.%
\cite{jiyeon}

\section{Results and Discussions}

\subsection{Group theory analysis and phonon assignment}

The 4$H$ structure has D$_{6h}$ symmetry,\cite{Donohue65,Donohue66} and four
molecular units in the primitive cell with eight Raman-active modes (2$%
A_{1g} $ + 2$E_{1g}$ + 4 $E_{2g}$), twelve IR-active modes (5$A_{2u}$ + 7$%
E_{1u}$) and eighteen silent optic modes ($A_{2g}$ + $A_{1u}$ + 3$B_{1g}$ + 2%
$B_{1u}$ + 5$B_{2u}$ + 6$E_{2g}$). The Raman-active modes are composed of $%
E_{2g}$ of Ba, $A_{1g}$, $E_{1g}$, and $E_{2g}$ of Ru, and $A_{1g}$, $E_{1g}$%
, and 2$E_{2g}$ of O. The point group for the 9$R$ structure is D$_{3d}$,%
\cite{Donohue65,Donohue66} with three molecular units in the primitive cell.
A factor group analysis predicts nine Raman-active modes (4$A_{1g}$ + 5$E_g$%
), sixteen IR-active modes (7$A_{2u}$ + 9$E_u$), and three silent optic
modes (2$A_{1u}$ + $A_{2g}$). The Raman-active modes are composed of \ $%
A_{1g}$ and $E_g$ of Ba, $A_{1g}$ and $E_g$ of Ru, and 2$A_{1g}$ and 3$E_g$
of O.

\begin{figure}[tbp]
\epsfig{file=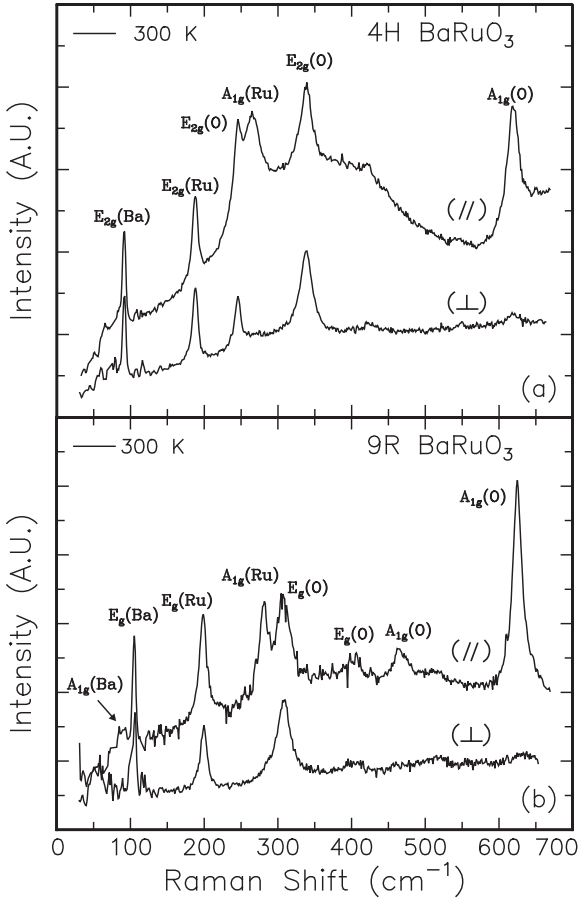,width=3.0in,clip=}
\vspace{2mm}
\caption{Polarized (//) and depolarized ($\perp $) Raman spectra of (a) 4$H$
and (b) 9$R$ BaRuO$_3$ in the $ab$-plane at 300 K.}
\label{4h-rt}
\end{figure}

It is noted that only the O ions participating in the face-sharing and only
the Ru ions in RuO$_6$ octahedra participating in the corner-sharing have
Raman-active phonon modes. O ions in both ruthenates are positioned at two
irreducible sites, i.e. a face-shared plane and an edge in RuO$_6$ blocks,
represented by O(1) and O(2) respectively, in Fig. 1. From the group
analysis, O(2) should have only the IR-active phonon modes ($A_{2u}$+2$E_{1u}
$ for 4$H$ BaRuO$_3$ and 2$A_{2u}$+3$E_u$ for 9$R$ BaRuO$_3$), while O(1)
should have both IR- and Raman-active modes. Unlike 4$H$ BaRuO$_3$, 9$R$
BaRuO$_3$ has two irreducible Ru-ion sites, i.e. a side and a center
position in the RuO$_6$ blocks, represented by Ru(1) and Ru(2) in Fig 1(b).
Similarly to the case of O ions, Ru(2) should have the IR-active phonon
modes ($A_{2u}$+$E_u$) only. So, it is expected that 4$H$ and 9$R$ BaRuO$_3$
show similar Raman-active phonon spectra in spite of the different layered
structures. For convenience, we will abbreviate O(1) and Ru(1) to O and Ru,
respectively.

Figure 2(a) shows the polarized and depolarized spectra of 4$H$ BaRuO$_3$.
Six phonon peaks are observed in the polarized spectra. In the depolarized
spectra, four peaks are observed at the same frequency positions with the
corresponding phonons in the polarized spectra. Note that both the $A_{1g}$
and the $E_{2g}$ modes in the D$_{6h}$ symmetry contribute to the polarized
signal, but only the $E_{2g}$ modes are present in the depolarized spectra,
which is an indicator for assigning the observed phonon modes.\cite{Cardona}
So, four modes observed in depolarized spectra can be assigned as $E_{2g}$
modes. Generally, the phonon frequencies related to the vibration of heavier
ions are lower than those of lighter ions. Thus, the four modes in ascending
order of their frequency are assigned as one $E_{2g}$(Ba), one $E_{2g}$(Ru)
and two $E_{2g}$(O). The rest of the modes in the polarized spectra are
assigned as $A_{1g}$(Ru) and $A_{1g}$(O) from the lowest frequency. Note
that according to the predictions of group theory, two $E_{1g}$ modes of Ru
and O cannot be observed in Raman spectra from the $ab$-plane of the thin
film sample.

The observed phonons in 9$R$ BaRuO$_3$ are assigned in a similar way. As
shown in Fig. 2(b), eight phonons are observed in the polarized spectra.
And, three clear modes, which are assigned as $E_g$(Ba), $E_g$(Ru) and $E_g$%
(O), are observed in the depolarized spectra. Though the mode near 403 cm$%
^{-1}$ in the polarized spectra is not clearly observed in the depolarized
spectra at 300\ K, a distinguishable feature of this mode is detected in the
depolarized spectra at low $T$. So, the mode is assigned as another $E_g$(O)
mode. The rest of the modes in ascending order of their frequency in the
polarized spectra are assigned as $A_{1g}$(Ba) near 85 cm$^{-1}$, $A_{1g}$%
(Ru) near 280 cm$^{-1}$, and two $A_{1g}$(O) near 463 cm$^{-1}$ and 625 cm$%
^{-1}$. The phonon assignments of 4$H$ and 9$R$ BaRuO$_3$ are summarized in
TABLE I.

On the other hand, Quilty {\it et al}.,\cite{Quilty93} from the $c$-axis
measurement of 9$R$ BaRuO$_3$, assigned the $A_{1g}$(Ru) mode at 280 cm$%
^{-1} $ as another O($E_g$) mode. We note, as we will discuss later, that $T$%
-dependence of this mode and its frequency are quite similar to those of the 
$A_{1g}$(Ru) mode in 4$H$ BaRuO$_3$. So, it is very likely that the mode at
280 cm$^{-1}$ is assigned as a $A_{1g}$(Ru) mode. It might be possible that
a weak $E_g$(O) and a strong $A_{1g}$(Ru) mode are at nearly the same
frequency.

\subsection{$T$-dependent phonon spectra}

Figures 3(a) and 3(b) show the $T$-dependent polarized spectra of 4$H$ and 9$%
R$ BaRuO$_3$, respectively. The spectra are shifted up for clarity. It is
observed that both compounds show similar phonon spectra. Except for the $%
E_{2g}$(O) mode near 245 cm$^{-1}$ in 4$H$ BaRuO$_3$, five distinct phonon
modes are observed at similar frequencies in both compounds. The $A_{1g}$
and $E_{2g}$ modes of 4$H$ BaRuO$_3$ correspond to the $A_{1g}$ and $E_g$
modes of 9$R$ BaRuO$_3$, respectively. It is noted that the corresponding
phonon modes in both ruthenates show similar $T$-dependent behavior.

\begin{figure}[tbp]
\epsfig{file=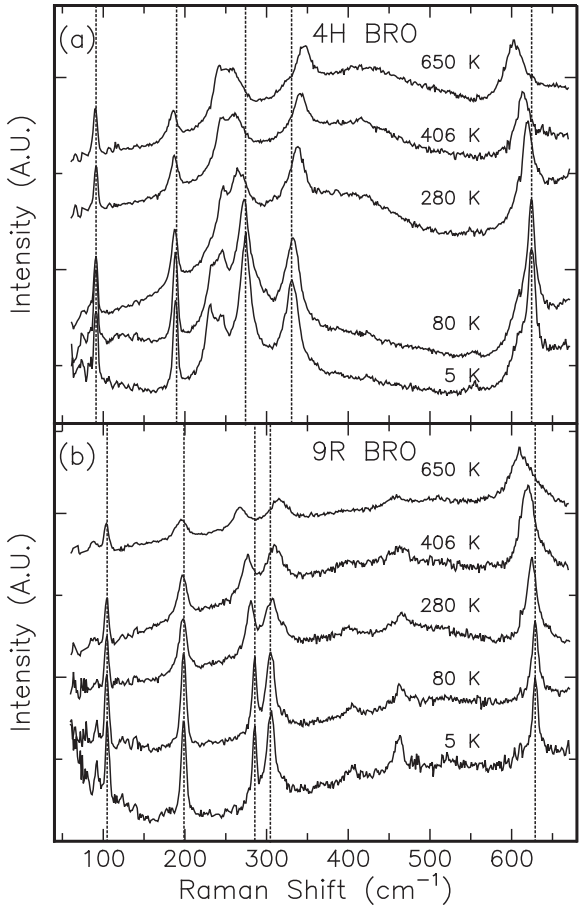,width=3.0in,clip=}
\vspace{2mm}
\caption{Temperature-dependent polarized spectra of (a) 4$H$ and (b) 9$R$
BaRuO$_3$ in the $ab$-plane. The spectra are shifted up for clear
presentation. The dotted lines are for guidance.}
\label{4h9r-t1}
\end{figure}

The $A_{1g}$ modes show different $T$-dependent behavior from the $E_{2g}$ ($%
E_g$) modes in both compounds. First, the $A_{1g}$(Ru) mode near 620 cm$%
^{-1} $ and the $A_{1g}$(O) mode near 270 cm$^{-1}$ strongly shift to higher
frequencies with decreasing $T$. [The $A_{1g}$(Ba) mode in 9$R$ BaRuO$_3$
shows a relatively weak hardening.] On the contrary, the $E_{2g}$(O) mode in
4$H$ BaRuO$_3$ near 340 cm$^{-1}$ and the $E_g$(O) mode in 9$R$ BaRuO$_3$
near 300 cm$^{-1}$ show a strong softening at lower $T$. These softenings
are quite unusual in that the general $T$-dependent behavior of phonons is
to show a hardening at lower $T$ due to an anharmonicity of lattice
vibrations. Even in a quite wide $T$ variation by $\sim $ 650 K, the $E_{2g}$%
(Ba, Ru) mode in 4$H$ BaRuO$_3$ and $E_g$(Ba, Ru) mode in 9$R$ BaRuO$_3$
show a very small change. On the other hand, as shown in Fig. 3(b), another $%
E_{2g}$(O) mode near 245 cm$^{-1}$, which is present only in the 4$H$ BaRuO$%
_3$, shows an anomalous $T$-dependence. This mode splits into two modes at
low $T$. [The details will be discussed in the next section.]

Figure 4 shows the detailed $T$-dependence of the phonons. The left panels,
Figs. 4 (a), (b), (c), and (d) show the $T$-dependences of the $A_{1g}$(O), $%
E_{2g}$(O), $A_{1g}$(Ru), and $E_{2g}$(Ru) modes, respectively, of 4$H$ BaRuO%
$_3$. The right panels, Figs. 4 (e), (f), (g), and (h) show the $T$%
-dependence of the phonons of 9$R$ BaRuO$_3$ corresponding to those of 4$H$
BaRuO$_3$, in sequence. While $A_{1g}$(O) modes show strong hardenings, $%
E_{2g}$(O) (or $E_g$) modes show softenings with decreasing $T$. Similarly, $%
A_{1g}$(Ru) modes show strong hardenings, but no significant change of $%
E_{2g}$(Ru) (or $E_g$) modes is observed. It is clear that $A_{1g}$ modes
show different $T$-dependent behavior from $E_{2g}$ ($E_g$) modes.\cite
{comment5}

\begin{figure}[tbp]
\epsfig{file=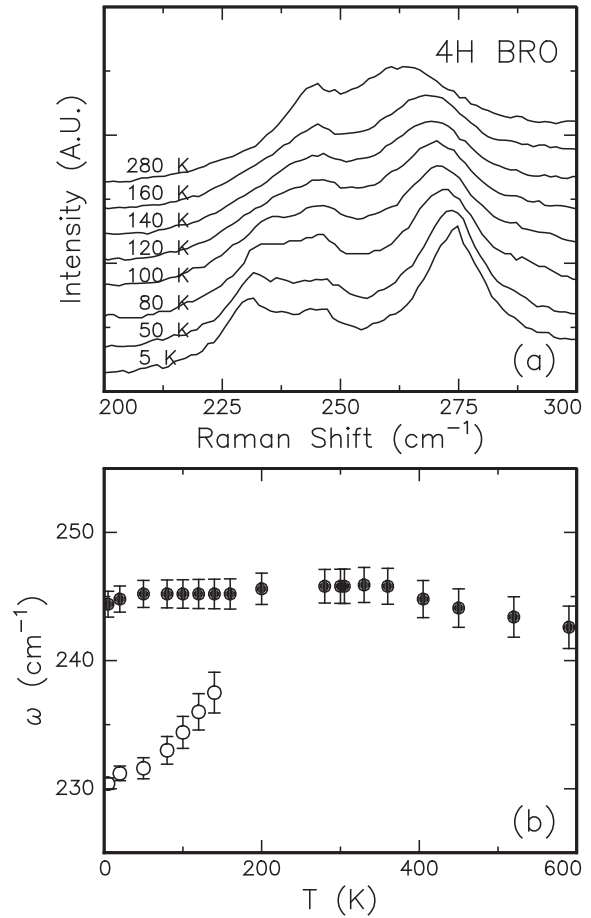,width=3.0in,clip=}
\vspace{2mm}
\caption{The temperature-dependence of the phonon frequencies of (a) $A_{1g}$%
(O), (b) $E_{2g}$(O), (c) $A_{1g}$(Ru), and (d) $E_{2g}$(Ru) in 4$H$ BaRuO$%
_3 $. (e) $A_{1g}$(O), (f) $E_g$(O), (g) $A_{1g}$(Ru), and (h) $E_g$(Ru) in 9%
$R$ BaRuO$_3$. }
\label{4h9r-t2}
\end{figure}

It is noted that the $T$-dependent behavior of the phonon modes closely
depends on the direction of lattice vibrations. As a tentative description, $%
E_g$ (or $E_{2g}$) and $A_{1g}$ modes are related to the vibrations in the $%
ab$-plane and along the $c$-axis, respectively. The different $T$-dependence
of the $A_{1g}$ and $E_{2g}$ (or $E_g$) modes could originate from their
anisotropic structural properties with the face-sharing of RuO$_6$ octahedra
along the $c$-axis, through which a strong anisotropic interaction, i.e. a
direct Ru metal-bonding, occurs.

The strong hardenings of $A_{1g}$ modes indicate that the bonding stiffness
along the $c$-axis becomes larger. This also implies that the interaction
along the $c$-axis, i.e. a direct Ru metal-bonding, becomes stronger. It is
noted that strong hardenings of $A_{1g}$ modes cannot be simply explained by
the variation of the $c$-axis lattice constant {\bf c}. The $A_{1g}$ modes
vibrating along the $c$-axis are expected to have a close relation with {\bf %
c}. However, while synchrotron XRD experiments with 4$H$ BaRuO$_3$ report
that its {\bf c} is changed just by $\sim $ 0.005 \AA\ with $T$ varying
between 30 K and 310 K,\cite{xrd-4h} the phonon frequency $\omega _{ph}$ of
the $A_{1g}$ modes is changed by $\sim $ 10 cm$^{-1}$ in the corresponding $T
$ range. These changes of the $\omega _{ph}$($A_{1g}$) are much larger than
those of the cuprates, where the stretching modes in Cu-O planes are quite
sensitive to Cu-O bonding distances; while the lattice constant in the $ab$%
-plane is reduced by about 0.1 \AA , the $\omega _{ph}$ increases higher by
100 cm$^{-1}$.\cite{Tajima91} The relatively large change of the $\omega
_{ph}$($A_{1g}$) in BaRuO$_3$ systems means that there must be another
electronic contribution to the strong hardening of the $A_{1g}$ modes in
addition to those of the lattice effect.\cite{khkim96} It is noted that
similar hardening of the phonon modes in some manganites are observed due to
charge ordering fluctuation.\cite{hjlee00} On the other hand, in the
infinite 1D chain cuprates, i.e. Ca$_{2-x}$Sr$_x$CuO$_3$, Drechsler {\it et
al}. suggested the possible existence of a dynamic Peierl-type distortion
and predicted phonon anomalies such as a hardening or a splitting.\cite
{Drechsler94} The strong hardenings of the $A_{1g}$ modes in the 1D-like
ruthenates might be related to the CDW fluctuation by the metal-bonding.

Unlike the $A_{1g}$ modes, $E_{2g}$ (or $E_g$) modes vibrating normally to
the metal bonding direction show softenings or no significant change.
Especially, strong softenings of the $E_{2g}$(O) (or $E_g$) modes are quite
unusual. It is noted that only the oxygens participating in the face-sharing
have Raman-active modes. So, the softenings of the $E_{2g}$(O) (or $E_g$)
modes indicate that the bonding stiffness among the face-shared oxygens
reduces and that there should be a kind of structural instability in the
face-shared O structure. This also implies that the Ru-O-Ru interaction
weakens as a Ru-Ru interaction strengthens, which could induce the stronger
1D-character. These differences in $T$-dependence between $A_{1g}$ and $E_g$
($E_{2g}$) modes in BaRuO$_3$ could be a unique feature reflecting a
structural instability due to the increase of the metal-bonding strength.%
\cite{comment2}

\subsection{Splitting of $E_{2g}$(O) mode in 4$H$ BaRuO$_3$}

Another important observation is that the $E_{2g}$(O) mode near 245 cm$^{-1}$
in 4$H$ BaRuO$_3$ shows a clear splitting. It is noted that this $E_{2g}$(O)
mode is not permitted in 9$R$ BaRuO$_3$ from group theory analysis. As shown
in Fig. 5(a), as $T$ decreases, the $E_{2g}$(O) mode becomes suppressed and
a new mode at lower frequency develops. Together with the softenings of
other $E_{2g}$(O) modes, this splitting clearly indicates the existence of a
structural instability in the face-shared structure. While the onset
temperature of the phonon splitting is not clear, as shown in Fig 5(b), some
anomaly is observed at $\sim $ 360 K, where the $T$-dependence of the $E_{2g}
$(O) mode is changed. This appears consistent with the optical observation
that the pseudogap feature is still observed at 300 K.\cite{yslee2} It is
very likely that this structural instability is closely related to the
origin of the pseudogap formation in the ruthenates, i.e. the CDW
instability.

\begin{figure}[tbp]
\epsfig{file=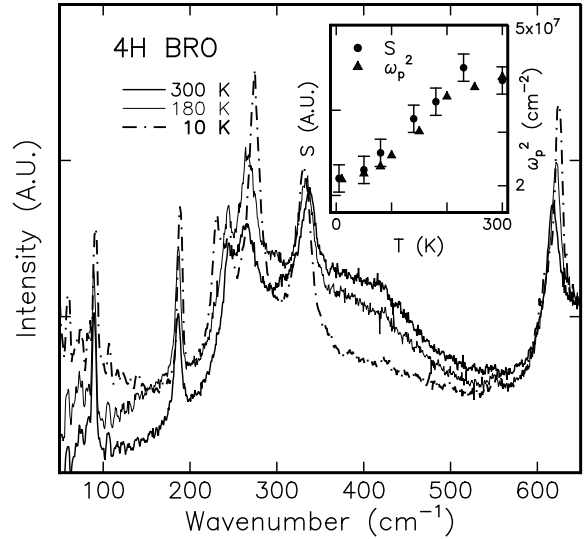,width=3.0in,clip=}
\vspace{2mm}
\caption{(a) Temperature-dependent behavior of $E_{2g}$(O) mode in 4$H$ BaRuO%
$_3$. (b) Phonon frequency vs. temperature. The solid and open circle
symbols represent the $E_{2g}$(O) mode and a new phonon mode, respectively.}
\label{4h-t1}
\end{figure}

It is noted that the phonon splitting happens in the $E_{2g}$(O) mode
vibrating normally to the metal-bonding direction. The splitting of an
IR-active phonon mode in the $ab$-plane was also observed for a static CDW 9$%
R$ BaIrO$_3$.\cite{Cao00} This implies that the charge modulation along the $%
c$-axis might be closely related to the structural instability in the $ab$%
-plane. On the other hand, it is interesting that only the $E_{2g}$(O) mode
near 245 cm$^{-1}$ shows the splitting. This means that the local structural
distortion related to this mode occurs, maintaining the total
crystallographic symmetry of the 4$H$ compound. More theoretical and
experimental studies are needed to understand the unusual CDW instability
and its detailed relation to the pseudogap formation.

\subsection{Electronic Raman continua}

Electronic Raman continua give important information about the electronic
excitation. Figure 6 shows $T$-dependent polarized spectra of 4$H$ BaRuO$_3$%
. As $T$ decreases, Raman continua below 200 cm$^{-1}$ increase. This
behavior is commonly observed in polarized and depolarized spectra in both
ruthenates. The increase of Raman continua in low frequency region could be
due to the reduction of the screening of free carriers on elastic
scatterings, such as Rayleigh responses. This is consistent with optical
observations that a reduction of carrier density $n$ occurs with pseudogap
formation.\cite{yslee1,yslee2} Similar behavior was observed in some
perovskite manganites with the metal-insulator transition.\cite{Yoon98}

\begin{figure}[tbp]
\epsfig{file=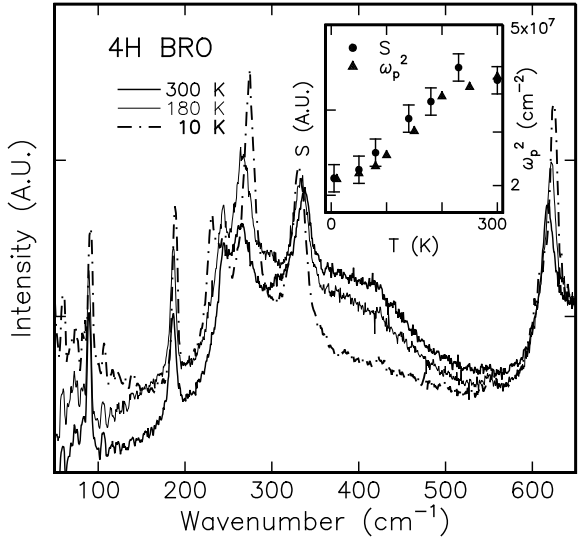,width=3.0in,clip=}
\vspace{2mm}
\caption{Temperature-dependent Raman spectra of the 4$H$ BaRuO$_3$ in the
polarized direction. Inset: the solid circle and the solid triangle symbols
represent the integrated Raman backgrounds $S$ and the square of the plasma
frequency $\omega _p^2$ (quoted from Ref. [2]).}
\label{4h-t2}
\end{figure}

Another important point is that a broad Raman continuum near 400 cm$^{-1}$
in 4$H$ BaRuO$_3$ is observed with a significant $T$-dependence.\cite
{comment4} As shown in Fig. 6, this broad continuum becomes suppressed with
decreasing $T$. Because the maximum of the continuum is at a relatively high
frequency, it cannot be an electronic Raman scattering by charge
fluctuations arising from electron-hole excitations near the Fermi energy.%
\cite{Klein84} And, its position is different from that of the pseudogap
position, $\sim $ 650 cm$^{-1}$.\cite{yslee2} To get qualitative physical
insights, we integrated the Raman backgrounds in the frequency region of 250 
$\sim $ 600 cm$^{-1}$, where the $T$-dependence is dominant.\cite{comment3}
Interestingly, as shown in the inset of Fig. 6, the integrated area $S$ is
strongly correlated with the square of the plasma frequency $\omega _p^2$,
obtained by optical measurement.\cite{yslee2} The reduction in $\omega _p^2$
originates mainly from a reduction in $n$ caused by a partial-gap opening on
the Fermi surface. The strong correlation between $S$ and $\omega _p^2$
indicates that the suppression of the distinct Raman excitation near 400 cm$%
^{-1}$ might be closely related to the pseudogap formation.

\section{Summary}

Raman spectra in 4$H$ and 9$R$ BaRuO$_3$ show interesting features related
to the metal-bonding formed through the face-sharing of RuO$_6$ octahedra.
The temperature-dependence of the observed phonons strongly depend on the
vibration direction with respect to the metal-bonding. For 4$H$ BaRuO$_3$,
another $E_{2g}$ mode of the oxygen participating in the face-sharing is
split clearly. The temperature-dependence of the broad electronic Raman
continua near 400 cm$^{-1}$ suggests a partial-gap opening on the Fermi
surface. These observations indicate that there occurs a kind of structural
instability due to the metal-bonding, which could be closely related to the
pseudogap formation in the BaRuO$_3$ systems.

\acknowledgments
We acknowledge Jaejun Yu at Seoul National University for his helpful
discussions. This work was supported by KOSEF through CSCMR. TWN
acknowledges Ministry of Science and Technology for financial supports
through the Creative Research Initiative program. EJO and ISY acknowledge
the support from the grant (No. KRF2000-015-DS0014) of Basic Science
Research Institute program of the Korea Research Foundation.

\newpage

\begin{table}[tbp]
\caption{Summaries of the Raman-active phonon modes in 4{\it H} and 9$R$
BaRuO$_3$ at 300 K.}
\begin{tabular}[t]{cccccc}
\multicolumn{3}{c}{4$H$ BaRuO$_3$} & \multicolumn{3}{c}{9$R$ BaRuO$_3$} \\ 
\tableline mode & frequency (cm$^{-1}$) & assignment & mode & frequency (cm$%
^{-1}$) & assignment \\ 
\tableline &  &  & $A_{1g}$ & 85 & Ba \\ 
$E_{2g}$ & 91 & Ba & $E_g$ & 105 & Ba \\ 
$E_{2g}$ & 187 & Ru & $E_{g}$ & 199 & Ru \\ 
$E_{2g}$ & 245 & O &  &  &  \\ 
$A_{1g}$ & 264 & Ru & $A_{1g}$ ($E_g$) & 282 & Ru (O) \\ 
$E_{2g}$ & 339 & O & $E_{g}$ & 307 & O \\ 
&  &  & $E_{g}$ & 403 & O \\ 
&  &  & $A_{1g}$ & 463 & O \\ 
$A_{1g}$ & 619 & O & $A_{1g}$ & 625 & O
\end{tabular}
\end{table}


\begin{references}
\bibitem{comment2}  The previous optical studies suggested that $T^{*}$
should be $\sim $ 430 K for 9$R$ BaRuO$_3$ and above 300 K for 4$H$ BaRuO$_3$%
. In our Raman spectra, weak anomalies in the $T$ region of 350 K $\sim $
450 K are observed in both BaRuO$_3$ materials.
\bibitem{comment4}  This electronic background is observed only in polarized
spectra. This indicates that the Raman excitation might be related to an
intraband transition within the conduction Ru 4$d$ band. This background is
hardly observed in 9$R$ BaRuO$_3$ with a smaller $n$ than 4$H$ BaRuO$_3$.
\bibitem{comment3}  It is assumed that the intensities of the phonons are
not changed with $T$ variation.
\end{references}
\end{document}